# Development of highly nonlinear polarization maintaining fibers with normal dispersion across entire transmission window

Dominik Dobrakowski[1,2], Anupamaa Rampur[1,2], Grzegorz Stępniewski[1,2], Alicja Anuszkiewicz[1,2], Jolanta Lisowska[1,2], Dariusz Pysz[1], Rafał Kasztelanic[1,2] and Mariusz Klimczak[1,2]

[1] Institute of Electronic Materials Technology, Glass Department, Wólczyńska 133, 01-919 Warsaw, Poland
[2] University of Warsaw, Faculty of Physics, Pasteura 5, 02-093 Warsaw, Poland

E-mail: mariusz.klimczak@itme.edu.pl

## Abstract

Determined polarization state of light is required in nonlinear optics applications related to ultrashort and single-cycle light pulse generation. Such short time scales require up to full octave of spectral width of light. Fiber-based, pulse-preserving and linearly polarized supercontinuum can meet these requirements. We report on the development – from linear simulations of the fiber structure, through fabrication of physical fibers to their versatile characterization – of polarization maintaining, highly nonlinear photonic crystal fibers, intended for femtosecond pumping at a wavelength of 1560 nm. Full octave of linearly polarized light around this wavelength would enable to cover amplification bandwidths of the three major fiber amplifiers from ytterbium doped systems up to thulium and holmium doped fiber amplifiers, with a coherent, linearly polarized seed signal. At the same time, an all-normal chromatic dispersion profile over an entire transmission window, and small dispersion of nonlinearity in the developed fibers, would facilitate use of commercially available femtosecond fiber lasers as pump sources for the developed fibers.

Keywords: Fibers; Birefringence; Normal dispersion; Nonlinear optics

PACS: fiber waveguides, 42.81.Qb, glasses, 42.70.Ce, nonlinear optics, 42.65.-k, frequency conversion (nonlinear optics), 42.65.Ky

## 1. Introduction

Coherence time of a light source is determined by the reciprocal of the spectral bandwidth. In the simplest view, the narrower the spectral line, the more coherent the light source is. Broad bandwidth light is usually incoherent. Mode-locked lasers, having bandwidths of tens of nanometers are nevertheless characterized by long coherence time. A femtosecond laser output pulse in the frequency domain consists of an equally spaced comb-like structure, pertaining to the coherent superposition of its resonator longitudinal modes [1]. The coherence time in this case stems from width of the "comb" lines (*e.g.* tens to hundreds of kHz) and not from the width of the envelope typically recorded as the laser spectrum [2].

Titanium-sapphire lasers have octave-spanning amplification bands, which enable obtaining few-cycle pulses directly from the oscillator. Mode-locked rare-earth doped fiber lasers offer far superior robustness, but their amplification bands span typically 50-250 nm in the near-infrared. The related limitations of few-cycle pulse generation with fiber lasers can be overcome by taking advantage of all-normal dispersion fiber-based supercontinuum generation. Octave spanning spectra in these conditions can be readily obtained under 50-100 fs pumping [3–5] and similar performance with commercially available, mode-locked fiber lasers operating at 1560 nm has also been reported [6].

The output supercontinuum pulses preserve the temporal profile and the comb-like structure of the driving lasers (albeit the pulses are dispersively stretched). In the case of 780 or 1030 nm femtosecond pumping from optical parametric chirped pulse amplification laser systems, these characteristics have been shown to enable temporal recompression to below 2 optical cycles, using active phase compensation [7]. A light pulse lasting single optical cycles has been demonstrated using erbium doped fiber amplifiers and advanced synthesis of spectrally broadened pulses due to dispersive wave generation and soliton self-frequency shift [8]. The best of the two approaches: preservation of the output pulse's temporal profile and fiber-based realization is possible to combine when a fiber amplifier is coherently seeded over its whole amplification band from a pulse-preserving supercontinuum source. Power scaling of such a system with a 1030 nm femtosecond laser front-end has been demonstrated up to a level compatible with attosecond pulse regime [9]. Coherent, full-bandwidth seeding of a thulium and holmium co-doped fiber amplifier has been recently discussed [6]. Importantly, use of a standard polarization-maintaining (PM) active fibers and fiber-optics components (pump-signal combiners or wavelength-division multiplexing – WDM elements) facilitates phase coherence and intensity stability during power (pulse energy) scaling, because beating among randomly polarized fiber modes is not present. This, however, requires the supercontinuum signal be linearly polarized. In-line fiber polarizers can be used, albeit spectral filtration of the seed is difficult to avoid. It is particularly important in case of $Tm^{3+}$+$Ho^{3+}$ doped fiber amplifiers, which possess the broadest bandwidth of all the three major fiber amplifiers.

Polarized and pulse-preserving supercontinuum can be obtained around 1030 nm with the use of commercially available fibers [10,11]. The available spectral width of supercontinuum obtainable in such a case does not cover $Tm^{3+}$+$Ho^{3+}$ amplification band. A PM microstructured silica fiber with highly Ge-doped core has been recently demonstrated, in which the dispersion at normal wavelengths has been engineered flat over near-infrared wavelengths up to around 2500 nm [12]. At this wavelength, a zero dispersion point was estimated with numerical simulations. Also, the effective area at around 2 µm wavelength was already reaching 40 µm$^2$, which is just about half of the area of commercial large mode area fibers [11].

We report, that it is possible to achieve birefringence of the weakly PM microstructured fibers in the order of 10$^{-4}$ reported earlier for silica all-normal dispersion (ANDi) photonic crystal fibers (PCFs) compatible with 1030 nm pumping [10], using relatively sturdy silicate soft glasses, to arrive at a PM ANDi fiber compatible with 1560 nm femtosecond lasers. A non-PM ANDi soft glass fiber reported recently in successful coherent supercontinuum application [6] is taken as a starting point in designing of two different PCF structures with intentional birefringence, all-normal dispersion profile across the fibers' transmission window and just over 10 µm$^2$ of effective area over the $Tm^{3+}$+$Ho^{3+}$ fiber amplification band. The presented discussion of results of numerical simulations of the fibers' properties is supported with detailed description and characterization of physically developed test fibers.

## 2. Structure design and material properties

We begin by revisiting of the starting structure, which was successfully demonstrated in coherent supercontinuum generation [4,13] and more recently in application at coherent seeding of ultrafast, $Tm^{3+}$+$Ho^{3+}$ fiber amplifier seeding [6]. The fiber structure is shown in a scanning electron microscope image (SEM) in figure 1. It has a hexagonal lattice, in which the air-holes are filled with a different type of glass rods at the preform assembly stage to form glass inclusions in the final fibers.

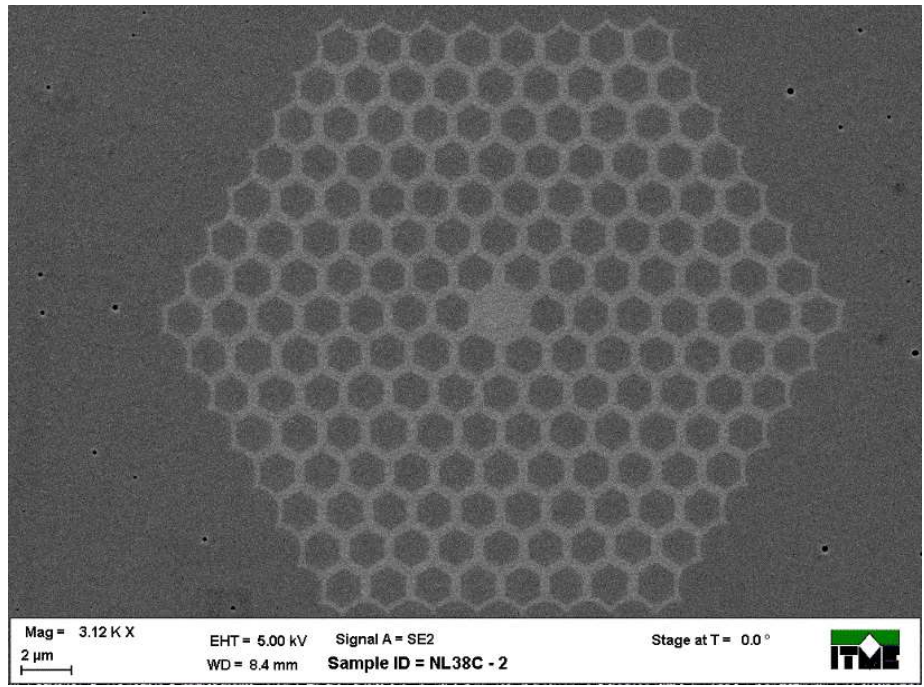

**Figure 1.** SEM image of the fabricated fiber structure, here referred as "base structure".

The structure in figure 1, earlier reported in [6,13] and recently also in [14], is specifically composed of Schott glasses labelled SF6 (core, lattice) and F2 (lattice filling, and surrounding tube). Material dispersion of both glasses is shown in figure 2.

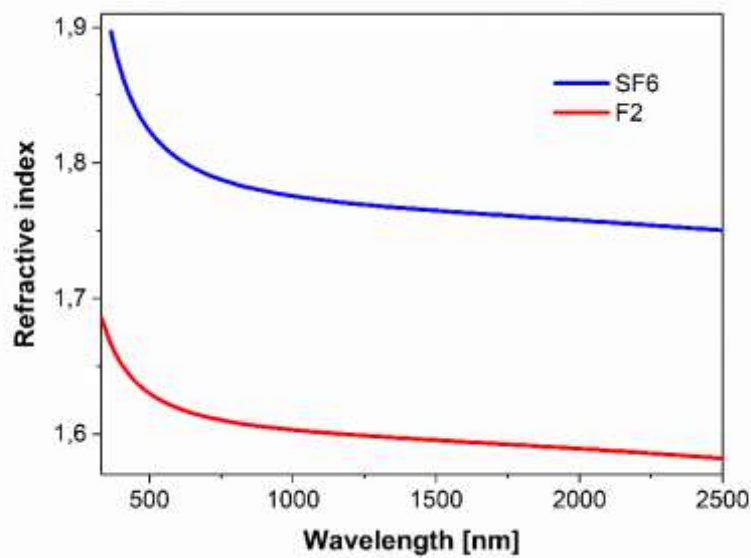

**Figure 2.** Material dispersion of the glass forming a core of the fiber (Schott SF6) and cladding with inclusions (Schott F2).

The refractive index of the core glass is higher than that of the core surrounding, therefore, the structure supports index guided propagation. Material dispersion was directly included in the simulation (initial values of material refractive index were wavelength-dependent, according to figure 2). This all-solid glass photonic crystal fiber approach facilitates engineering of flattened normal dispersion profiles, because chromatic dispersion of the fiber is then manipulated by both the waveguide (lattice topology) and the material (two glass types) contributions. One of such structures is taken as the starting point in the design and fabrication of a polarization maintaining variant.

A PCF with this type of lattice can be fabricated using the common stack-and-draw procedure. Two ways of introducing birefringence to a hexagonal PCF lattice are possible. Stress elements can be introduced in the fiber structure like in the classic Panda or bow-tie fibers – in the PCF domain and specifically in context of coherent supercontinuum generation PCFs, this has been used in such PCFs compatible with 1030 nm femtosecond pumping [10]. As a result of the built-in stress, a change of the refractive index in the structure's transverse plane is introduced. High birefringence of the fiber can also be achieved using the side-hole approach, where the mode confined in the core experiences defined optical axes with dramatically different refractive indices, due to presence of air-holes at two opposite sides of the core [15]. This approach is impractical with a type of structure shown in figure 1, because of the intricate role of the hexagonal lattice in shaping of the dispersion and mode confinement properties [16]. An elliptical deformation of the fiber structure, *i.e.* its squeeze along one of the axes, has therefore been investigated as a possible alternative means of inducing birefringence.

The proposed bow-tie modification of the original base fiber design is shown in figure 3(a). Compared to the base structure in figure 1, this design has the number of photonic lattice rings decreased by one (6 instead of 7). The lattice constant is Λ = 1.73 µm, width of the hexagonal inclusion is d = 1.35 µm, and d/Λ = 0.77. The outer diameter of the whole designed structure is 125 µm. The structure consists of three types of glasses, differing of refractive index values and mechanical properties. Two of them are again Schott SF6 and F2 glasses, and the third one is a modified F2 glass, labelled F2/05, with slightly changed refractive index ($n_D$ = 1.617) and significantly increased thermal expansion coefficient (table 1). The composition of the glass was as following (compound, mol.%): $SiO_2$ - 70.17, PbO - 15.55, $Na_2O$ - 8.08, $K_2O$ - 6.07, $As_2O_3$ - 0.13. Other parameters of the glasses, necessary for mechanical simulation were considered as temperature-dependent and equal to: Young's modulus - $5.547 \cdot 10^{10}$ / $5.763 \cdot 10^{10}$ Pa, Poisson's ratio - 0.238 / 0.236 and density – 5180 / 3610 kg/m$^3$ in room temperature (293.15 K) for SF6 and F2 glass, respectively.

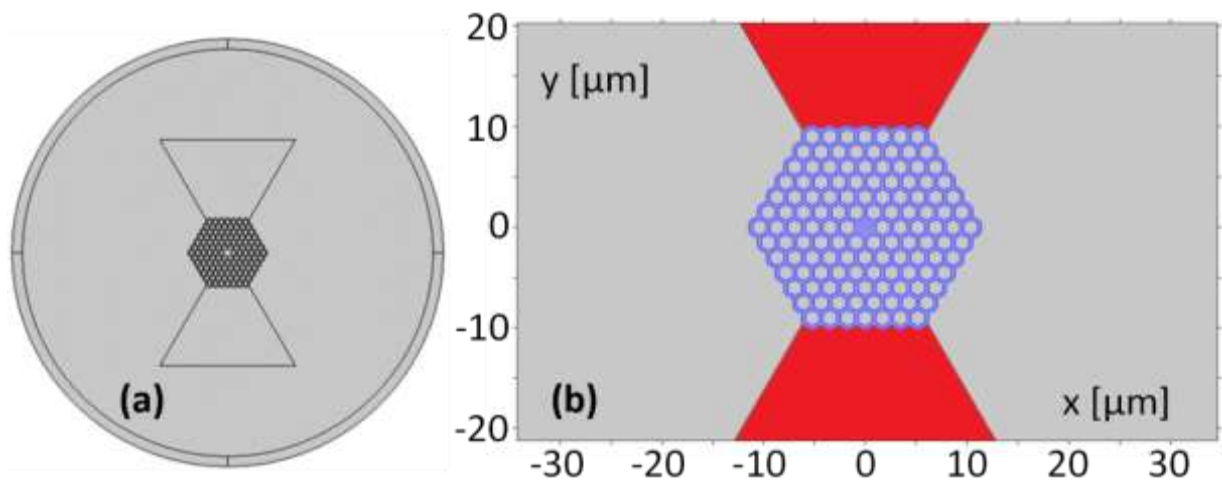

**Figure 3.** (a) Photonic crystal fiber design with stress-inducing components. (b) Magnified central part of the designed fiber with colored areas indicating different used glasses, blue for Schott SF6 glass, red for F2/05 glass (stress rods) and grey for Schott F2.

In order to take into account the contribution of stress to the change of the refractive index in the structure, an approach similar to reported in [17] was applied. Numerical simulations were performed using the finite element method, operationalized in COMSOL Multiphysics software. The simulation process was done in two steps. In the first one, the *Structural Mechanics* module was used and the simulation involved induced stress calculations, while the second step was an optical-domain simulation with the use of the *Wave Optics* module.

The chosen mesh density was adapted to the size of the structure's element, starting from 8.38 µm of the maximal element size and 0.0375 µm for the minimal one (fiber's cladding), to 0.15/0.05 µm (microstructure). The boundary condition was *Perfectly Matched Layer*, having a thickness of 3 µm, located at the external boundary of the cladding.

The main assumption considering the stress induced in the fiber structure is, that the stress builds into the structure at temperatures below the glass transition temperature $T_g$ [17]. The $T_g$ values, as well as thermal expansion coefficients for each of the glasses used for design of the fiber, are summarized in table 1.

**Table 1.** Glass transition temperatures and thermal expansion coefficients of glasses used as fiber structure materials.

| Glass | Glass transition temperature ($T_g$) [°C] | Thermal expansion coefficient ($\alpha_{20,\ 300°C}$) [1/K] | Ref. |
|---|---|---|---|
| Schott SF6 | 423 | $9 \cdot 10^{-6}$ | [18] |
| Schott F2 | 431 | $9.092 \cdot 10^{-6}$ | This work |
| F2/05 | 417.6 | $10.55 \cdot 10^{-6}$ | This work |

It is then assumed, that under the built-in stress, the x and y components of a refractive index tensor are changed. The refractive index tensor is defined as:

$$n = \begin{bmatrix} n_x & 0 & 0 \\ 0 & n_y & 0 \\ 0 & 0 & n_z \end{bmatrix} \quad (1)$$

The change of its components are calculated according to a formula:

$$n_i = n_0 - C \cdot \sigma_j \quad (2)$$

where *i* and *j* denote *x* and *y* components, *C* is the stress-optical coefficient, $\sigma$ is the appropriate stress vector component and $n_0$ is the value of refractive index without the influence of mechanical stress.

It is further assumed, that z component of the refractive index tensor does not change. The value of stress-optical coefficient depends on the material. In the simulation, they are assumed equal to $0.65 \cdot 10^{-12}$ m$^2$/N for SF6 glass [18] and $2.81 \cdot 10^{-12}$ m$^2$/N for F2 [18] and F2/05. The calculated x component of the stress tensor and birefringence induced in the fiber structure (difference between y and x components of refractive index tensor) are presented in figure 4.

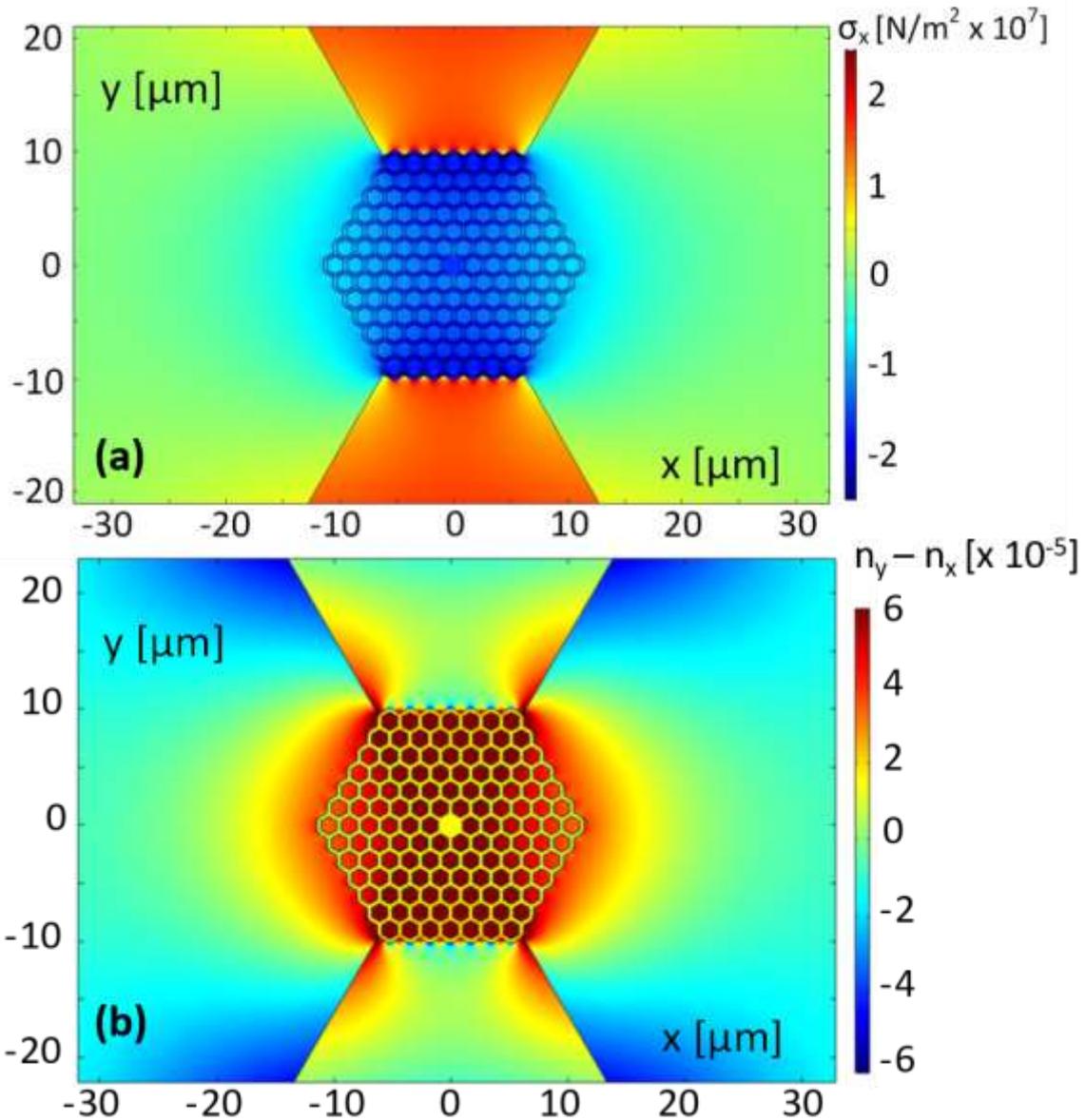

**Figure 4.** (a) x component of tensor of stress, (b) stress induced birefringence in the fiber with stress rods.

The obtained difference between both components of the stress tensor has a moderate value, at the order of $10^{-5}$. This can be assigned to the fact, that the stress reaching the core area is largely diminished by the photonic lattice – which in turn is crucial for dispersion shaping and mode confinement. The effective mode area ($A_{eff}$) of the fundamental mode was also calculated in the simulations and it is within range of 2.5 to 15 µm² over a wavelength span of 900 to 2750 nm, as shown in figure 5. In particular, the wavelength dependence of the effective area, which translates into dispersion of nonlinearity [19] is desirable when weak – is such a case nonlinear response similar to that at the pump wavelength would facilitate efficient spectral broadening at the red-shifted edge of the propagating pulse's spectrum.

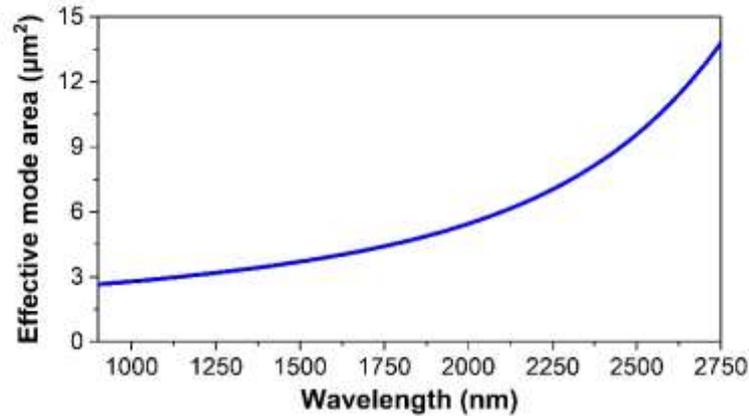

**Figure 5.** Calculated effective mode area of a fundamental mode of the simulated fiber with stress rods.

## 3.     Fabricated structures and dispersion characterization

The preform for a test structure of the ANDi PCF with stress-inducing rods, discussed in the previous section, was stacked with the use of rods and capillaries by the standard stack-and-draw technique. Physically developed test structure has slight ellipticity, as shown in figure 6, in a SEM image. This ellipticity was caused by different softening temperatures for both glasses of the fiber cladding (the one of the stress rods and of the glass filling). The difference in temperatures results from modified thermal expansion coefficients of the stress rods material (table 1).

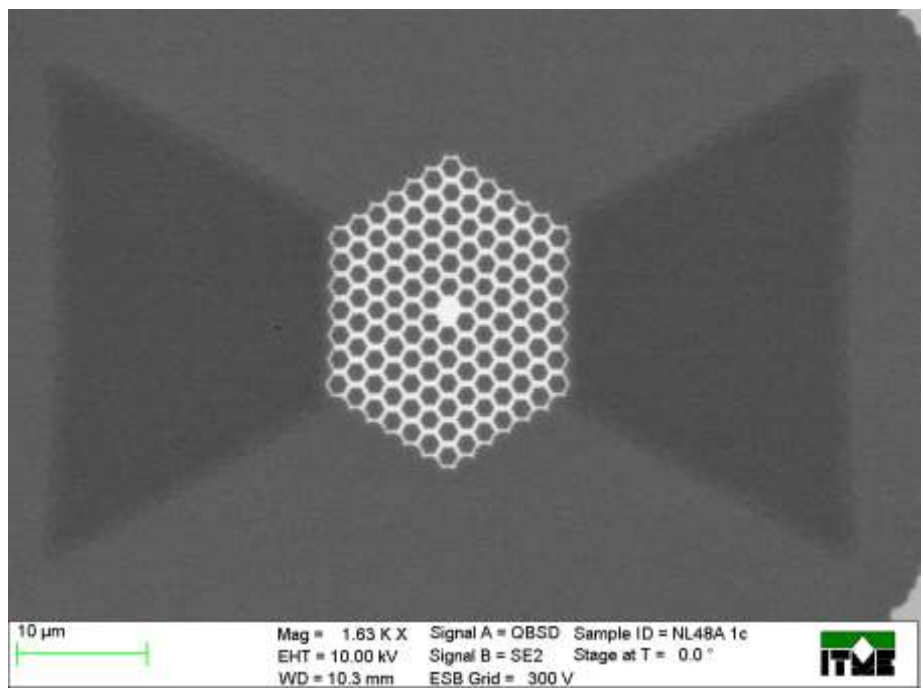

**Figure 6.** SEM image of the fabricated fiber structure with stress-inducing components ( "bow-tie" structure).

The second proposed way of introducing birefringence to the hexagonal photonic crystal base structure is making the lattice elliptical. The number of lattice rings in the photonic cladding of the physically developed fiber was reduced to 5, and air holes with diameter comparable to the size of the photonic lattice structure, were placed on each side of the lattice at the preform stacking stage. The air holes were collapsed during drawing in order to introduce ellipticity of the fiber microstructure. The fabricated elliptically-deformed fiber, is shown in figure 7.

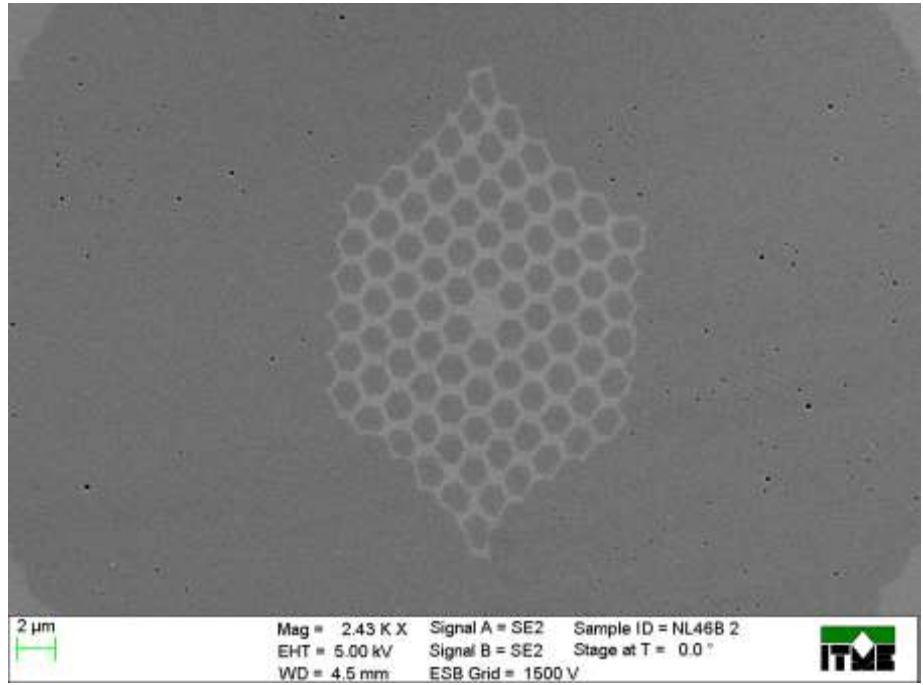

**Figure 7.** SEM image of the fabricated fiber structure with elliptical deformation.

Dimensions of the structure's core were equal to 2.68 x 1.91 µm, which means, that the structure was more squeezed, than the discussed "bow-tie" fiber. Lattice widths at diagonals were of 25.8/16.3 µm and the lattice constant was 1.99 µm.

The fabricated test fibers were then characterized experimentally, and specifically chromatic dispersion and birefringence measurements were performed. The developed test fibers were intended to validate technological feasibility of fabricating ANDi PM fibers with reasonable value of birefringence, simultaneously with maintained all-normal dispersion profile, across the transmission window. The fibers were not optimized with respect to loss and broadband attenuation measurements were not performed. Based on typical coupling efficiencies achievable with similar fibers, e.g. [4,13,14,16] and mean power output from the fibers reported in this work, we estimated loss level of around 6 dB/m. This is improvable in further iterations of the technological process of their fabrication. We note that this unoptimized and considerable attenuation level did not preclude recording of clear interferograms in dispersion and birefringence measurements with use of fiber sample lengths typical for femtosecond supercontinuum generation. Chromatic dispersion in the developed fiber was measured for each of the polarization axes using an unbalanced Mach-Zehnder interferometer setup, schematically shown in figure 8. The method was explained with details in [20].

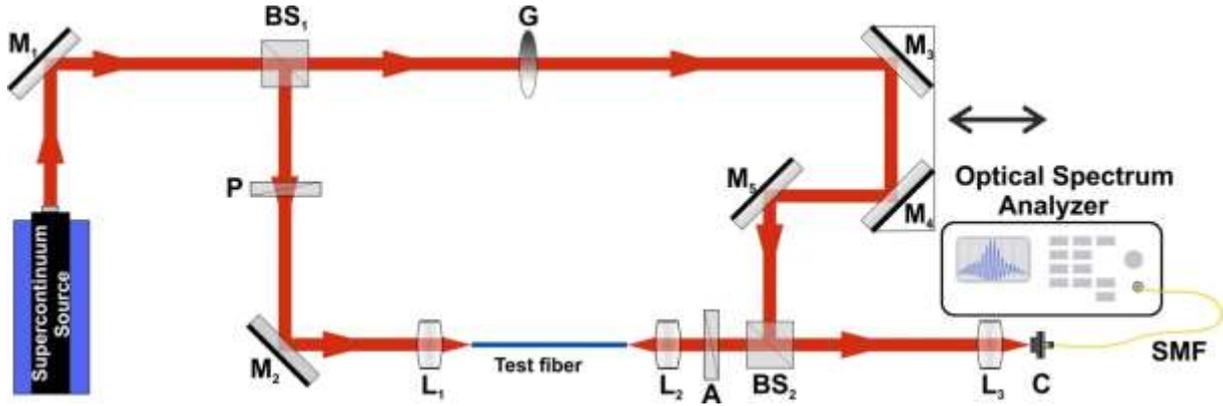

**Figure 8.** Mach-Zehnder interferometer setup for dispersion measurement: M – silver mirrors, P – polarizer, BS – beam-splitter, G – gradient neutral-density filter, L – lenses, A – polarization analyzer, C - coupler, SMF – single mode fiber.

A collimated supercontinuum (450 - 2400 nm, Leukos SM-30-450) is used as the light source. The light is divided into the two arms of the interferometer with a beam splitter with 1-6 μm anti-reflection coating ($BS_1$). Neatly cleaved test fiber is placed in the test arm of the interferometer. A 60× microscope objective ($L_1$) was used to couple the supercontinuum light into the test fiber, and an aspheric lens ($L_2$) at the other end of the fiber collimated the output light. Coupling of light into the fiber was monitored by imaging the output end of the fiber onto an InGaAs camera, which also confirmed excitation of the fundamental mode in the investigated fiber in all measurements. The reference arm has a gradient filter (G) to match the equal intensity in both arms. The two mirrors ($M_3$ and $M_4$) on a linear translation stage in the reference arm allow to compensate the optical path length in the measurement arm. A broadband linear polarizer was placed in the measurement arm before and after the fiber as shown in figure 8. The polarizer and analyzer are rotated carefully to excite only one, either fast or slow axis, at a time. The geometry of the setup and length of the linear stages enables working with fibers of length from 10 to 20 cm. Typically used fiber length in our measurement was around 14 cm. Light from both arms is combined in a beam splitter ($BS_2$) and collimated. Lens ($L_3$) and coupler (C) is used to collect the light from the interferometer, which is then transferred through a single mode fiber (SMF) to an optical spectrum analyzer (OSA), which records from 1200 nm to 2400 nm (Yokogawa AQ6375). Length of the reference arm is adjusted to determine an optical path difference between the test and reference arms. The characteristic wide interference fringes over a wide spectral range are recorded to determine the chromatic dispersion of the fiber. Form this the relative compensation lengths Δx(λ) of the reference arm were collected for different equalization wavelength positions. This data was used to obtain the relative group index $N_{rel}(\lambda)$:

$$N_{rel}(\lambda) = \frac{2 \cdot \Delta x(\lambda)}{L} + 1 \qquad (3)$$

where L is the length of the fiber.

Measured $N_{rel}(\lambda)$ was fitted with 5$^{th}$ order polynomial, that is Cauchy dispersion equation as follows:

$$N_{rel}(\lambda) = A_1\lambda^{-4} + A_2\lambda^{-2} + A_3 + A_4\lambda^2 + A_5\lambda^4 \qquad (4)$$

From this we can obtain A1-A5 parameters to further calculate the chromatic dispersion as:

$$D(\lambda) = \frac{1}{c}\frac{dN_{rel}(\lambda)}{d\lambda} = \frac{1}{c}(-4A_1\lambda^{-4} - 2A_2\lambda^{-2} + 2A_4\lambda^2 + 4A_5\lambda^4) \qquad (5)$$

The data obtained during measurements are fitted with equation $N_{rel}(\lambda)$ as mentioned in (4). $N_x$ and $N_y$ represents the relative group index for X and Y polarizations respectively for "bow-tie" as shown in figure 9. and for elliptical fiber as shown in figure 10. The gap between the measurements in figure 9 is due to water absorption and noise structure made it difficult to measure the equalization wavelength precisely.

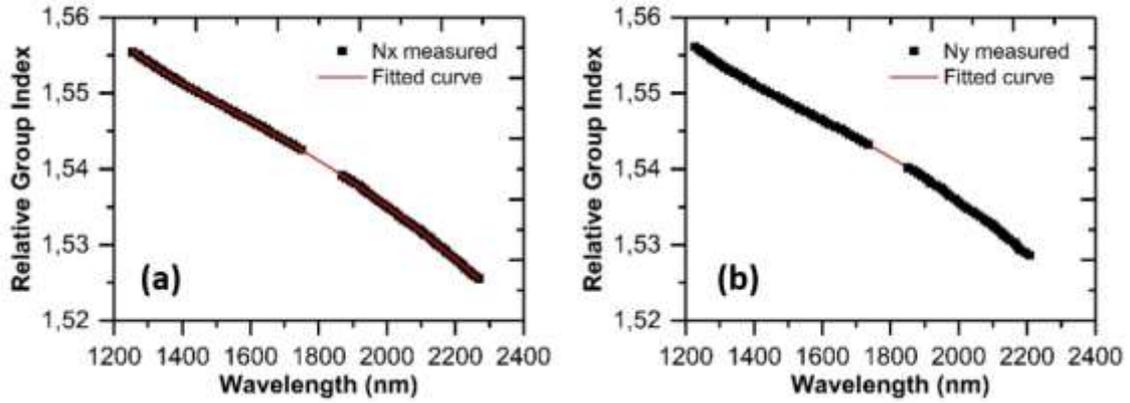

**Figure 9.** Relative group index obtained from measurements for "bow-tie" fiber (a) X polarization and (b) Y polarization. Solid dots are measured points and red curve was obtained by fitting.

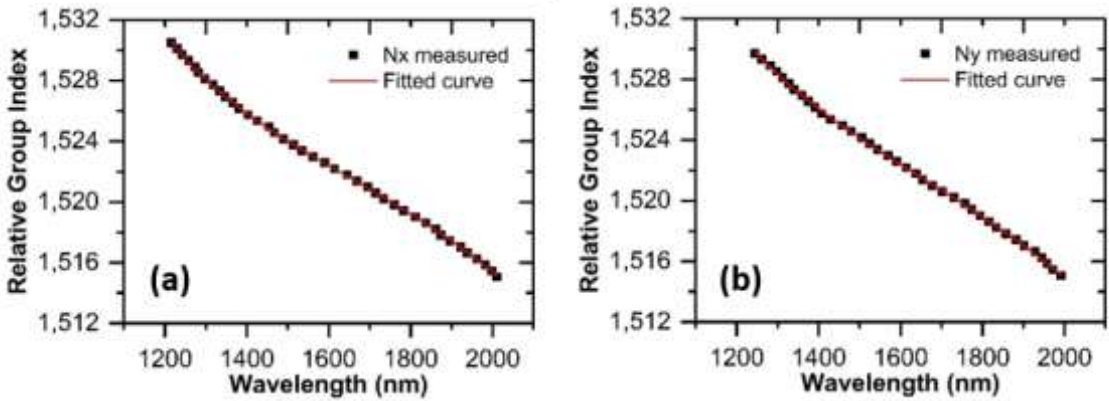

**Figure 10.** Relative group index obtained from measurements for elliptical fiber (a) X polarization and (b) Y polarization. Solid dots are measured points and red curve was obtained by fitting.

Fitting parameters of $N_{rel}(\lambda)$ obtained for both polarization modes in the elliptical and "bow-tie" fibers are presented in table 2. From these parameter chromatic dispersion (D) is calculated from equation (5).

**Table 2.** Fitting parameters for reproducing the measured dispersion parameter D.

| Fiber | $A_1$ | $A_2$ | $A_3$ | $A_4$ | $A_5$ |
|---|---|---|---|---|---|
| Bow-tie ($E_x$ pol) | 4.83E-02 | -5.21E-02 | 1.58 | -0.01009 | 2.81E-05 |
| Bow-tie ($E_y$ pol) | 3.39E-02 | -2.71E-02 | 1.57 | -0.0053 | -4.07E-04 |
| Elliptical ($E_x$ pol) | 8.19E-03 | 2.34E-02 | 1.51 | 0.0058 | -1.20E-03 |
| Elliptical ($E_y$ pol) | -2.36E-03 | 3.54E-02 | 1.50 | 0.00574 | -1.16E-03 |

To obtain D in ps/nm/km, express the speed of light c in km/ps and the wavelength λ in μm.

Numerical simulation was done, so that the structure would resemble the physically developed one. The fabricated structures were simulated with the assumption, that the squeeze is only in one direction. The geometrical parameters of the "bow-tie" fiber were taken from real structure: the core size was equal to 2.56 x 1.94 µm, lattice widths were of 24.5/18.7 µm, average lattice constant of 1.68 µm and stress applying zones of the size of 38.6 x 21.5 µm. Measured dispersion profiles of the fiber with stress-inducing components, compared with numerical results, are presented in figure 11.

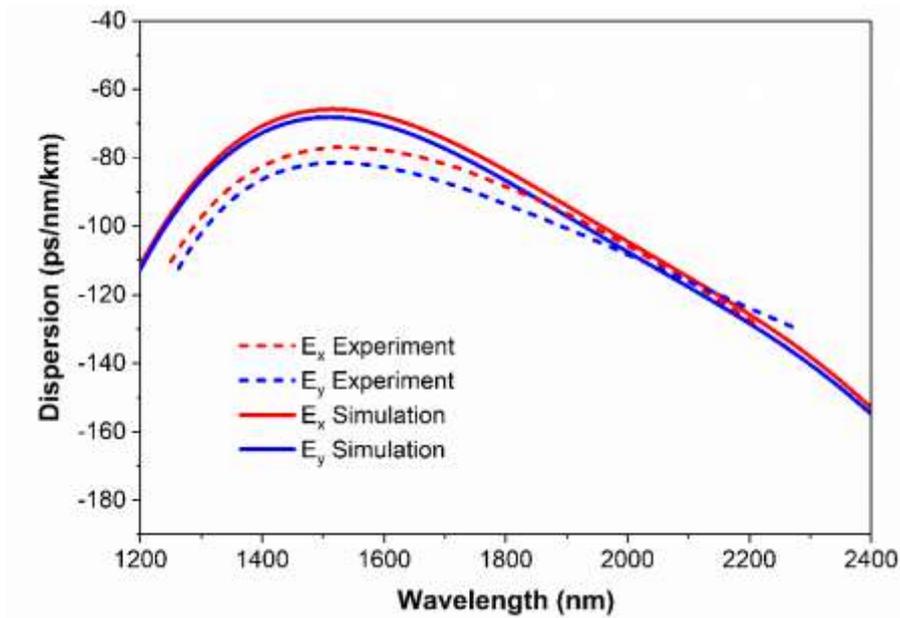

**Figure 11.** Comparison between the calculated (solid line) and experimental (dashed line) dispersion of the "bow-tie" fiber in the range of 1200 to 2400 nm.

Dispersion remains at normal values in the whole considered wavelength range and is in a very good agreement with dispersion profiles obtained numerically. A small difference between the numerical simulation and the measurement can be explained by diffusion during drawing, not taken into account in simulations.

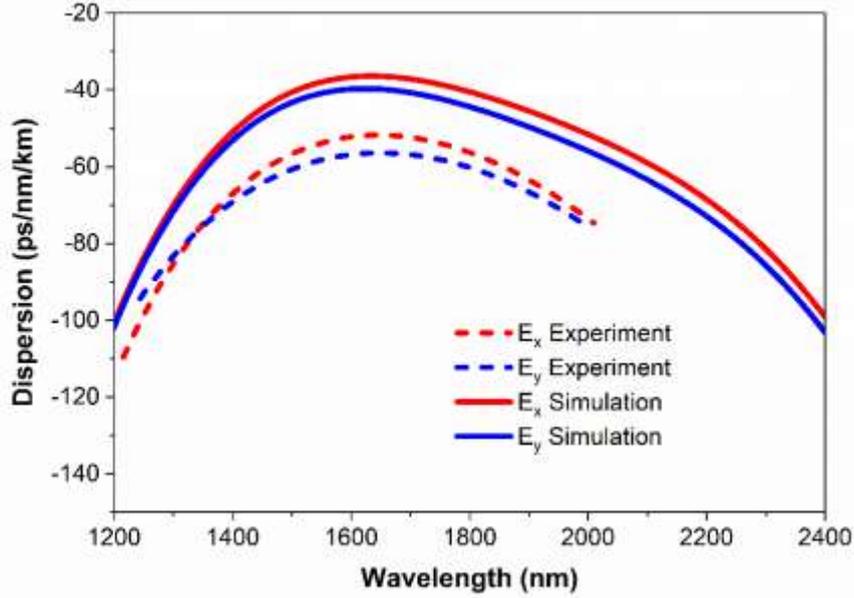

**Figure 12.** Comparison between the calculated (solid line) and experimental (dashed line) dispersion of elliptical fiber in the range of 1200 to 2400 nm.

Dispersion of the elliptical structure was calculated using the same numerical approach as for the other considered lattice layouts. The results of these calculations performed for both polarization axes of the fiber are shown and compared with physically measured dispersion profiles in figure 12. The experiments allowed to confirm, that the dispersion remains normal in the whole considered spectral range and has a very similar profile compared to the fabricated "bow-tie" structure, as well (dispersion characteristic shown in figure 11).

## 4. Birefringence characterization

In highly birefringent fiber the two orthogonal polarization modes experience different propagation constants $\beta_x(\lambda)$ and $\beta_y(\lambda)$ due to slightly varying refractive indices $n_x$ and $n_y$, respectively. When the light is launched into the fiber and one of the fiber's polarization modes is excited, the coupling between the two modes is greatly reduced and state of polarization is maintained along the fiber. The magnitude of birefringence is characterized by the beat length $L_B(\lambda)$ of two polarization modes, which is defined as:

$$L_B(\lambda) = \frac{2\pi}{\beta_x(\lambda) - \beta_y(\lambda)} \quad (6)$$

then the phase birefringence $B(\lambda)$ can be expressed by:

$$B(\lambda) = n_x - n_y = \frac{\lambda}{L_B(\lambda)} \quad (7)$$

and the group birefringence $G(\lambda)$ as:

$$G = B(\lambda) - \lambda \frac{dB(\lambda)}{d\lambda}. \quad (8)$$

The phase and group birefringence characteristics obtained numerically for the modelled structure, shown in figure 3, are presented in figure 13.

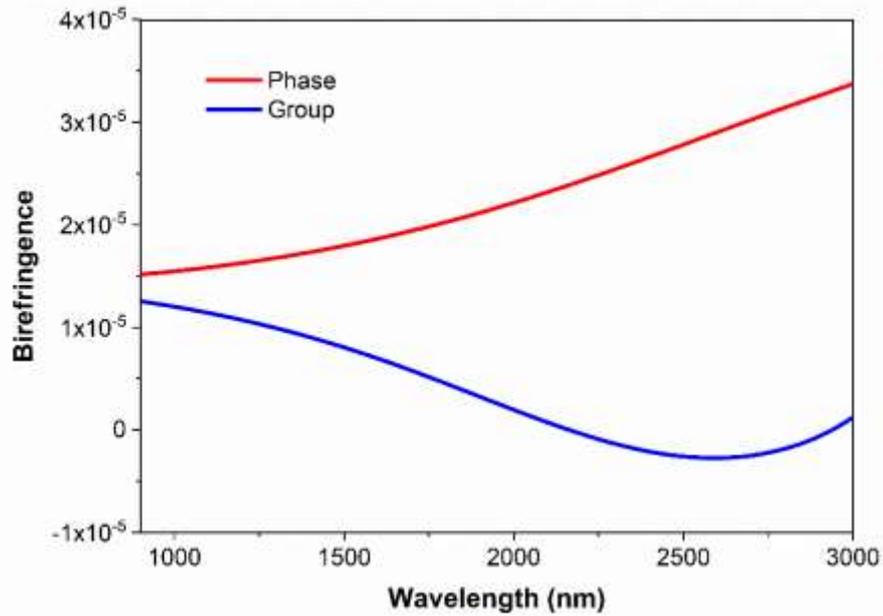

**Figure 13.** Calculated phase and group birefringence of the simulated "bow-tie" fiber.

The obtained values of phase birefringence are relatively low in the order of $10^{-5}$, *i.e.* the same order of magnitude as an induced birefringence in a material, see figure 4(b). However, similar birefringence values were observed in [21] as sufficient to improve quality of an all-normal dispersion supercontinuum pulse in a fiber with nominally smaller, unintentional form birefringence. Birefringence in the structure as shown in figure 4 originates from built-in stress. The group birefringence takes lower absolute values than the phase birefringence (figure 13), and at wavelengths close to the material transmission cutoff of the fiber glass, it takes values of an opposite sign.

Group birefringence can be measured using a simple experimental setup, which uses the standard crossed polarizer method as shown in figure 14.

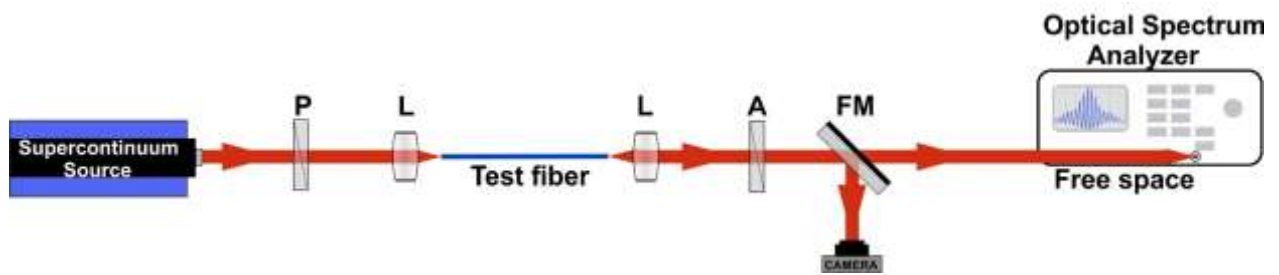

**Figure 14.** Experimental setup for group birefringence measurement P – polarizer, L – lenses, A – polarization analyzer, FM – flip mirror.

Similar to the dispersion measurement, a supercontinuum was used as a light source, and the light was coupled into the fiber with a 60× microscope objective. The output light from the fiber was coupled into the OSA over free space, after collimating with an aspheric lens. The coupling efficiency into the fiber and the mode structure was monitored using IR camera and a flip mirror. Broadband linear polarizer at the input was adjusted in such a way to excite both polarization modes in the fiber and analyzer at the output was oriented to obtain the interference birefringence pattern. Group birefringence (G) is obtained from the recorded interferogram using expression:

$$|G| = \frac{\lambda_0^2}{\Delta\lambda \cdot L} \tag{9}$$

where $\lambda_0$ and $\Delta\lambda$ are the average wavelength and the distance between the two successive fringes, respectively, and L is the length of the fiber.

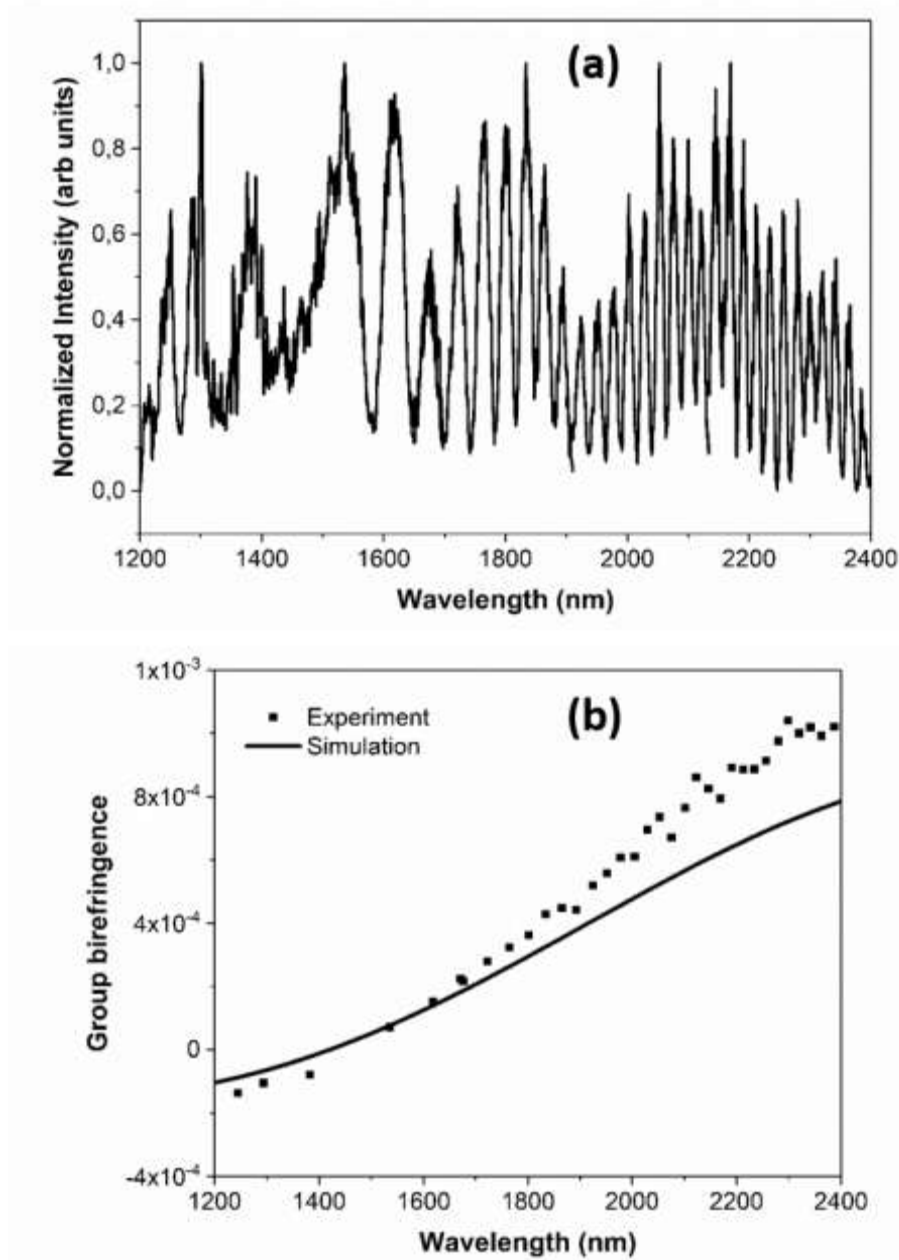

**Figure 15.** (a) Interference fringes recorded in group birefringence measurement. (b) Group birefringence comparison between simulated (solid) and measured (dots) profiles obtained for the "bow-tie" fiber.

Figure 15(a) presents interference fringes obtained for group birefringence of the "bow-tie" fiber. Coupling to the fiber was adjusted to obtain best possible fringes at different wavelengths. Intensity in y axis is normalized and fringes from different spectral regions were stitched. The wide fringe around 1400 nm corresponds to the zero group birefringence wavelength. Figure 15(b) shows that group birefringence obtained experimentally is in very close agreement with results of numerical simulations. The sign of group birefringence cannot be determined experimentally, so it was established by comparing with the profile obtained from numerical simulations.

Group birefringence of the "bow-tie" test fiber takes much larger values than estimated numerically for an ideal (not squeezed) structure shown in figure 3. For the longest wavelengths recorded experimentally it reaches the order of $10^{-3}$. We note that this large difference of group birefringence between the designed structure and the physical fiber is assigned to the ellipticity of the latter.

The birefringence characteristic measured for the elliptical fiber, shown in figure 16(b), is at much higher values, than for the regular structure with stress-inducing elements - numerical results shown in figure 6 - and also slightly higher than those of the "bow-tie" structure with both ellipticity and stress rods. In our elliptical fiber, measurements and simulations revealed, that group birefringence changes sign around roughly 1500 nm and increases with wavelength up the order of $10^{-3}$ at about 2000 nm, as shown in figure 16(b).

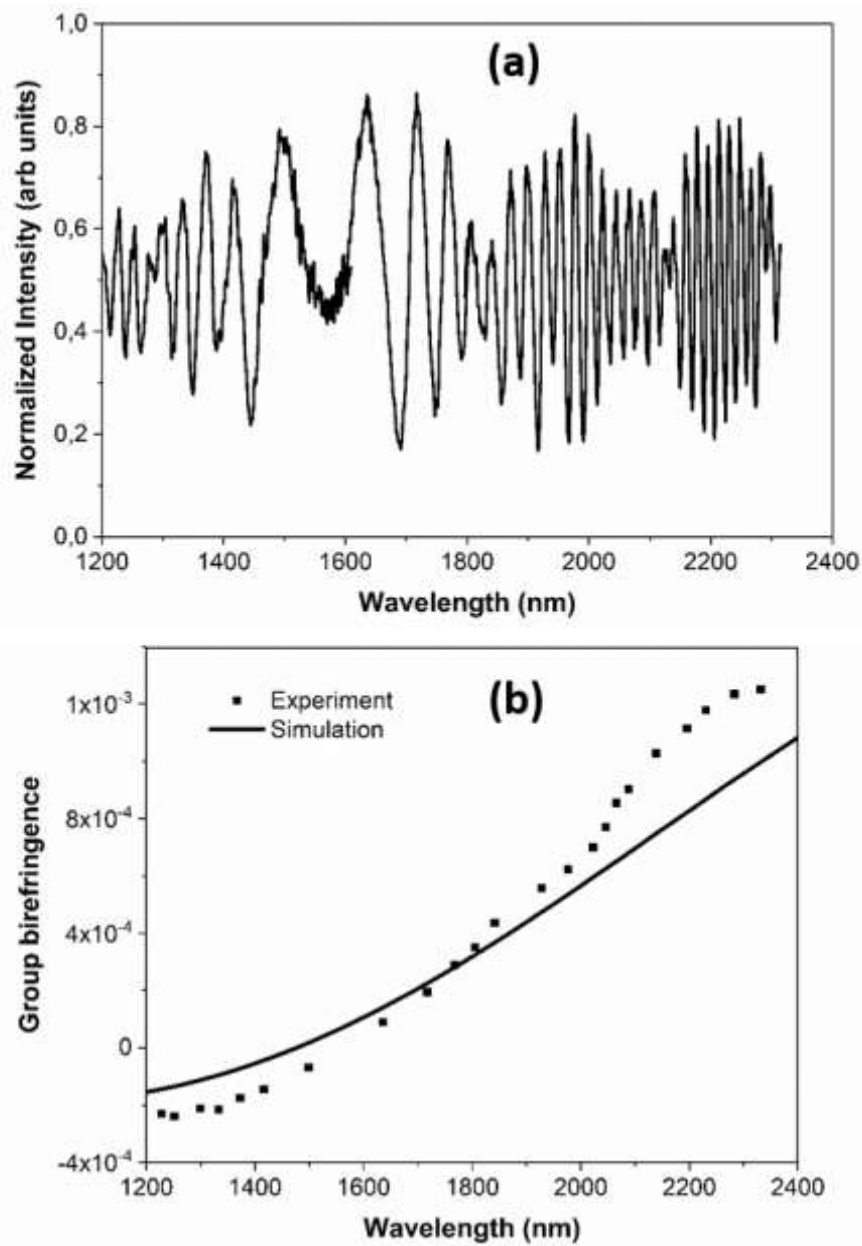

**Figure 16.** (a) Interference fringes recorded in group birefringence measurement of elliptical fiber. (b) Group birefringence comparison between simulated (solid) and measured (dots) profiles obtained for elliptical fiber.

Additionally, by performing numerical simulations, a possibility of introducing a comparable birefringence in a structure with stress rods and without ellipticity, was examined. For this purpose, a simulation of the structure presented in figure 3 was done with increased value of thermal expansion coefficient of the glass forming the stress rods by a factor of 1.5. The rest of the parameters was the same, as in the initial simulation. The resulting birefringence is shown in figure 17.

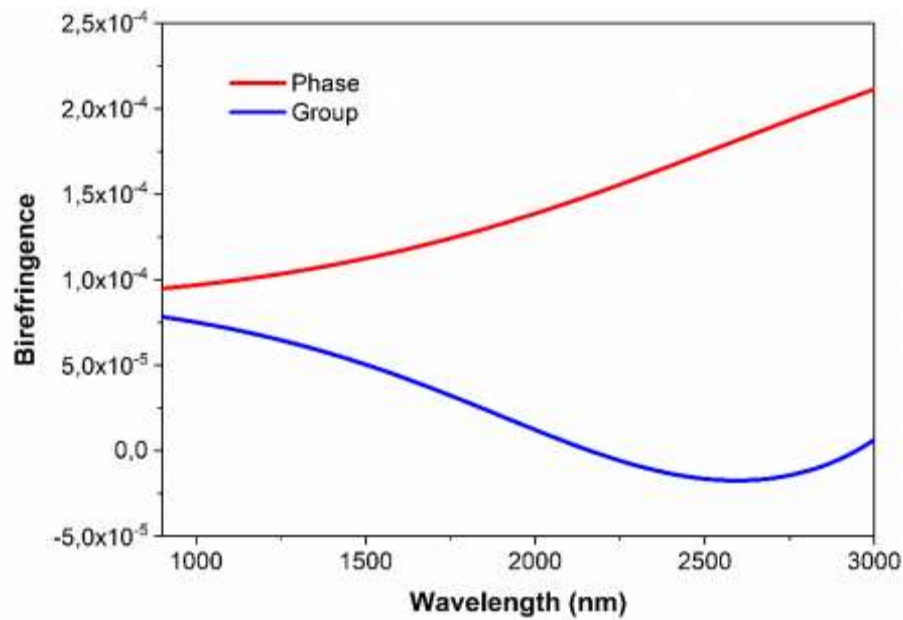

**Figure 17.** Calculated birefringence of a modified simulated "bow-tie" fiber.

Enhancing the thermal expansion coefficient of the glass of stress applying zones allows for increasing of the mechanical stresses introduced into the fiber, which in turn results in a significant rise of the birefringence (figure 17). The obtained values are approximately one order of magnitude greater, than in the case of the initial structure (figure 13). This proves a possibility of obtaining a highly birefringent PCF with a regular structure, but requiring a substantial improvement of one of the stress applying zones' material properties.

## 5. Discussion

A summary of parameters of two soft glass, weakly-PM, ANDi fibers investigated here, in comparison with literature results on birefringence in all-normal dispersion photonic crystal fibers is presented in table 3.

**Table 3.** Birefringence results obtained in all-normal dispersion photonic crystal fibers for wavelength ranges of the three rare-earth doped fiber-based femtosecond lasers.

| No. | Fiber (type, material, way of introducing birefringence) | ca. 1000 nm ($Yb^{3+}$ fs laser) | ca. 1560 nm ($Er^{3+}$ fs laser) | ca. 2000 nm ($Tm^{3+} + Ho^{3+}$ fs laser) | Reference |
|---|---|---|---|---|---|
| 1 | ANDi PCF optimized for 1000 nm, silica glass, bow-tie | $4.2 \cdot 10^{-4}$ | - | - | [10] |
| 2 | ANDi PCF opt. for 1500-2000 nm, Ge:silica, stress rods | - | $1.4 \cdot 10^{-4}$ | $1.7 \cdot 10^{-4}$ | [12] |
| 3 | Ultra-high numerical aperture PM | - | $1.7 \cdot 10^{-5}$ | $9 \cdot 10^{-5}$ | This work, |

| | fiber opt. for 1900 – 2100 nm (Coherent-Nufern PM2000D) | | (group) | (group) | [22] |
|---|---|---|---|---|---|
| 4 | ANDi PCF opt. for 1500-2000 nm, silicate soft glass, elliptical | - | 0 (group) | 6.6·10$^{-4}$ (group) | This work (elliptical) |
| 5 | ANDi PCF opt. for 1500-2000 nm, silicate soft glass, bow-tie | - | 1·10$^{-4}$ (group) | 6.1·10$^{-4}$ (group) | This work ("bow-tie") |

ANDi PM fibers on the soft glass platform are vital to extending the dispersion engineering freedom for highly nonlinear fibers for polarized, coherent supercontinuum generation. The fiber designs proposed here, have experimentally confirmed birefringence properties corresponding to existing designs realized and reported for the silica glass fiber technology. Although the fiber structures presented here are weakly PM, they are characterized by significantly enhanced nonlinear properties. This includes higher nonlinearity related to high value of nonlinear refractive index of the SF6 glass used for the core and lattice of the discussed fibers. It is however fostered with weak dependence of the effective mode area on wavelength, which stems from the waveguide design. This in turn results in strong nonlinear response over the Tm$^{3+}$ and Ho$^{3+}$ amplification wavelengths (1900-2200 nm) and holds the promise of the nonlinear soft glass platform PCFs in reproducing or even extending excellent nonlinear performance at these wavelengths, previously demonstrated with silica ANDi PCFs at the Yb$^{3+}$ band (1000-1100 nm). Comparison of relevant parameters of selected fibers is given in table 4.

**Table 4.** Nonlinear properties of all-normal dispersion photonic crystal fibers at near-infrared wavelengths.

| No. | Fiber (type, material, way of introducing birefringence) | n$_2$ [x10$^{-20}$ m$^2$/W] | A$_{eff}$ [μm$^2$] | Estimated nonlinear coefficient (γ) [1/W/km] | Ref. |
|---|---|---|---|---|---|
| 1 | ANDi PCF optimized for 1000 nm, silica glass, bow-tie | 2.68 | 3.8 (1064 nm) | ~37 (1064 nm) | [23,24] |
| 2 | ANDi PCF opt. for 1500-2000 nm, Ge:silica, stress rods | 2.6 | 20.2 (2000 nm) | 4 (2000 nm) | [12] |
| 3 | Ultra-high numerical aperture PM fiber opt. for 1900 – 2100 nm (Coherent-Nufern PM2000D) | 2.68 | 12.6 (2000 nm) | 6.7 (2000 nm) | [23,25] |
| 4 | ANDi PCF opt. for 1500-2000 nm, silicate soft glass, elliptical | 21 | 5.4 (2000 nm) | 122.2 (2000 nm) | This work (elliptical), [13] |
| 5 | ANDi PCF opt. for 1500-2000 nm, silicate soft glass, bow-tie | 21 | 5.8 (2000 nm) | 113.7 (2000 nm) | This work ("bow-tie"), [13] |

The values of estimated nonlinear coefficient in table 4, apart from fiber 1 (value explicitly given in [23]), were calculated using the formula:

$$\gamma = \frac{2\pi n_2}{\lambda A_{eff}}, \quad (10)$$

and the parameters taken from sources referred in table 4.

With the effective mode area at the level of 3.7 μm$^2$, corresponding to nonlinear refractive index (n$_2$) of 21·10$^{-20}$ m$^2$/W (at wavelength of 1500 nm) [13], the soft glass fibers reported here constitute an alternative to silica-based fibers, because of the potential to decrease requirements on the peak pump power for efficient operation in the nonlinear regime. At wavelength of 2 μm, they allow for nonlinear coefficient even of the order of $10^2$ $W^{-1}km^{-1}$ (table 4).

## 6. Conclusions

Numerical simulations and characterization of physically developed test structures of all-solid glass, all-normal dispersion, weakly PM photonic crystal fibers were performed. The obtained results indicate feasibility of delivering PM fibers with engineered, normal dispersion characteristics and nonlinearity earlier reported for non-PM, nonlinear PCFs for octave-spanning, pulse preserving supercontinuum generation. Two types of birefringence introduction were investigated. A hexagonal, all-solid glass photonic lattice of a non-PM fiber, reported earlier in [13] was taken as a starting point. Microstructure layouts with a "bow-tie" type stress rods around the photonic lattice and with core ellipticity were then investigated for birefringence properties. By introducing "bow-tie" stress rods using glass with thermal expansion coefficient of $10.55 \cdot 10^{-6}$ 1/K, weak birefringence at the order of $10^{-5}$ was estimated with numerical simulations. Fabricated bow-tie test fiber was squeezed during drawing and the measured birefringence reached around $10^{-3}$ at a wavelength of 2000 nm. This was confirmed with numerical simulations performed for geometrical parameters of this test fiber. These simulations further revealed, that in this particular fiber, the microstructure ellipticity counteracted the effect of the stress rods, nullifying it entirely and introducing birefringence of its own. The dominant effect of core ellipticity in birefringence of the proposed fibers was further confirmed by numerical simulations and measurements on a physical fiber featuring only core ellipticity without any stress rods. Numerical simulations allowed to estimate, that a stress-induced birefringence of around $10^{-4}$ in a structure without any ellipticity would be possible, if the stress rods glass had thermal expansion coefficient a factor of 1.5 larger, than that of the glass used in the actual technological process.

The developed fibers confirm feasibility of obtaining weakly-PM functionality, while preserving a normal dispersion characteristic over wavelengths matching the mature and emerging optical fiber amplifiers based on $Yb^{3+}$, $Er^{3+}$ and $Tm^{3+}/Ho^{3+}$ dopants. At the same time, the nonlinear parameters of the demonstrated test fibers are significantly enhanced over normal dispersion, weakly-PM microstructured fibers realized in the silica glass technology. Thus, the developed fibers are an interesting alternative for applications such as seed signal source generators for ultrafast fiber amplifiers reported recently using a $Tm^{3+}+Ho^{3+}$ fiber amplifier and a non-polarized, femtosecond-pumped ANDi supercontinuum seed [6]. Further work with the developed fibers is planned and will be focused on detailed, spectro-temporal characterization of femtosecond-pumped supercontinuum generation in the PM regime.

### Acknowledgements

This work has been carried out in scope of the project *High temporal quality ultrashort pulse generation for coherent seeding of high power near- and mid-infrared optical amplifiers* realized within the First TEAM programme of the Foundation for Polish Science co-financed by the European Union under the European Regional Development Fund (project number First TEAM/2016-1/1).

### References:

[1]   Hänsch T W 2006 Nobel Lecture: Passion for precision *Rev. Mod. Phys.* **78** 1297–309

[2]   Wei D, Takahashi S, Takamasu K and Matsumoto H 2009 Analysis of the temporal coherence function of a femtosecond optical frequency comb *Opt. Express* **17** 7011–8

[3]  Heidt A M, Hartung A, Bosman G W, Krok P, Rohwer E G, Schwoerer H and Bartelt H 2011 Coherent octave spanning near-infrared and visible supercontinuum generation in all-normal dispersion photonic crystal fibers *Opt. Express* **19** 3775–87

[4]  Klimczak M, Siwicki B, Skibiński P, Pysz D, Stępień R, Heidt A, Radzewicz C and Buczyński R 2014 Coherent supercontinuum generation up to 2.3 µm in all-solid soft-glass photonic crystal fibers with flat all-normal dispersion *Opt. Express* **22** 18824–18824

[5]  Tarnowski K, Martynkien T, Mergo P, Poturaj K, Soboń G and Urbańczyk W 2016 Coherent supercontinuum generation up to 2.2 µm in an all-normal dispersion microstructured silica fiber *Opt. Express* **24** 30523–36

[6]  Hodasi J M, Heidt A, Klimczak M, Siwicki B and Feurer T 2017 Femtosecond seeding of a Tm-Ho fiber amplifier by a broadband coherent supercontinuum pulse from an all-solid all-normal photonic crystal fiber Optics InfoBase Conference Papers vol Part F82-CLEO_Europe 2017

[7]  Heidt A M, Rothhardt J, Hartung A, Bartelt H, Rohwer E G, Limpert J and Tünnermann A 2011 High quality sub-two cycle pulses from compression of supercontinuum generated in all-normal dispersion photonic crystal fiber *Opt. Express* **19** 13873–9

[8]  Krauss G, Lohss S, Hanke T, Sell A, Eggert S, Huber R and Leitenstorfer A 2010 Synthesis of a single cycle of light with compact erbium-doped fibre technology *Nature Photon.* **4** 33–6

[9]  Rothhardt J, Demmler S, Hädrich S, Limpert J and Tünnermann A 2012 Octave-spanning OPCPA system delivering CEP-stable few-cycle pulses and 22 W of average power at 1 MHz repetition rate *Opt. Express* **20** 10870–8

[10]  Liu Y, Zhao Y, Lyngsø J, You S, Wilson W L, Tu H and Boppart S A 2015 Suppressing Short-Term Polarization Noise and Related Spectral Decoherence in All-Normal Dispersion Fiber Supercontinuum Generation *J. Light. Technol.* **33** 1814–20

[11]  NKT Photonics 2018 *Large Mode Area Photonic Crystal Fiber specification* (online) available: https://www.nktphotonics.com/lasers-fibers/product/large-mode-area-photonic-crystal-fibers/ (Accessed 27 04 2018)

[12]  Tarnowski K, Martynkien T, Mergo P, Poturaj K, Anuszkiewicz A, Béjot P, Billard F, Faucher O, Kibler B and Urbanczyk W 2017 Polarized all-normal dispersion supercontinuum reaching 25 µm generated in a birefringent microstructured silica fiber *Opt. Express* **25** 27452–27452

[13]  Klimczak M, Siwicki B, Zhou B, Bache M, Pysz D, Bang O and Buczyński R 2016 Coherent supercontinuum bandwidth limitations under femtosecond pumping at 2 µm in all-solid soft glass photonic crystal fibers *Opt. Express* **24** 29406--29416

[14]  Klimczak M, Siwicki B, Heidt A and Buczyński R 2017 Coherent supercontinuum generation in soft glass photonic crystal fibers *Photonics Research* **5** 710–27

[15]  Anuszkiewicz A, Martynkien T, Mergo P, Makara M and Urbanczyk W 2013 Sensing and transmission characteristics of a rocking filter fabricated in a side-hole fiber with zero group birefringence *Opt. Express* **21** 12657–67

[16]  Martynkien T, Pysz D, Stępień R and Buczyński R 2014 All-solid microstructured fiber with flat normal chromatic dispersion *Opt. Lett.* **39** 2342–5

[17] Tarnowski K, Anuszkiewicz A, Poturaj K, Mergo P and Urbanczyk W 2014 Birefringent optical fiber with dispersive orientation of polarization axes *Opt. Express* **22** 25347–53

[18] Schott 2017 *Optical Glass Data Sheet* (online) available: http://www.schott.com/d/advanced_optics/ac85c64c-60a0-4113-a9df-23ee1be20428/1.3/schott-optical-glass-collection-datasheets-english-17012017.pdf (Accessed 27 04 2018)

[19] Kibler B, Dudley J M and Coen S 2005 Supercontinuum generation and nonlinear pulse propagation in photonic crystal fiber: influence of the frequency-dependent effective mode area *Applied Physics B* **81** 337–42

[20] Hlubina P, Kadulová M and Ciprian D 2012 Spectral interferometry-based chromatic dispersion measurement of fibre including the zero-dispersion wavelength *Journal of the European Optical Society: Rapid Publications* **7**

[21] Tu H, Liu Y, Liu X, Turchinovich D, Lægsgaard J and Boppart S A 2012 Nonlinear polarization dynamics in a weakly birefringent all-normal dispersion photonic crystal fiber : toward a practical coherent fiber supercontinuum laser *Opt. Express* **20** 1113–28

[22] Ciąćka P, Rampur A, Heidt A and Klimczak M 2018 Dispersion Measurement of Ultra - High Numerical Aperture Fibers covering Thulium , Holmium , and Erbium Emission Wavelengths *J. Opt. Soc. Am. B* **35** 1301–7

[23] Traore A, Lalanne E and Johnson A M 2015 Determination of the nonlinear refractive index of multimode silica fiber with a dual-line ultrashort pulse laser source by using the induced grating autocorrelation technique *Opt. Express* **23** 17127–37

[24] Ainnotech 2018 *Nonlinear Photonic Crystal Fiber specification* (online) available: http://ainnotech.com/ainnotech/pdf/04/1_3/6NKT_NL-1050-NEG-1_Nonlinear%20Photonic%20Crystal%20Fiber.pdf (Accessed 27 04 2018)

[25] Nufern 2018 *Optical Fiber specification* (online) available: http://www.nufern.com/pam/optical_fibers/3017/PM2000D/ (Accessed 27 04 2018)